# Reconstructing shock front of unstable detonations based on multi-layer perceptron

Lin Zhou [1,2], Honghui Teng [1], Hoi Dick Ng [3], Pengfei Yang [4,5,*], Zonglin Jiang [4,5]

[1] School of Aerospace Engineering, Beijing Institute of Technology, Beijing 100081, China;
[2] State Key Laboratory of Laser Propulsion & Application, Beijing Power Machinery Institute, Beijing 100074, China;
[3] Department of Mechanical, Industrial and Aerospace Engineering, Concordia University, Montreal, QC, H3G 1M8, Canada
[4] State key laboratory of high temperature gas dynamics, Institute of Mechanics, Chinses Academy of Sciences, Beijing, 100190, China
[5] School of Engineering Sciences, University of Chinese Academy of Sciences, Beijing 100049, China

*Corresponding author. Email: young1505@foxmail.com (Pengfei Yang)

**Abstract**

The dynamics of frontal and transverse shocks in gaseous detonation waves is a complex phenomenon bringing many difficulties to both numerical and experimental research. Advanced laser-optical visualization of detonation structure may provide certain information of its reactive front, but the corresponding lead shock needs to be reconstructed building the complete flow field. Using the Multi-Layer Perceptron (MLP) approach, we propose in this study a shock front reconstruction method which can predict evolution of the lead shock wavefront from the state of the reactive front. The method is verified through the numerical results of one- and two-dimensional unstable detonations based on the reactive Euler equations with a one-step irreversible chemical reaction model. Results show that the accuracy of the proposed method depends on the activation energy of the reactive mixture, which influences prominently the cellular detonation instability and hence, the distortion of the lead shock surface. To select the input variables for training and evaluate their influence on the effectiveness of the proposed method, five groups, one with six variables and the other with four variables, are tested and analyzed in the MLP model. The trained MLP is tested in the cases with different activation energies, demonstrates the inspiring generalization capability. This paper offers a universal framework for predicting detonation frontal evolution and provides a novel way to interpret numerical and experimental results of detonation waves.

# 1. Introduction

A detonation is a supersonic, combustion-driven compression wave across which there is a significant pressure and temperature increase. Due to its destructive nature and rapid release of energy, a wealth of fundamental research on this subject can be found in the literature and has been of wide interest for terrestrial and astrophysical explosions [1–4]. In recent years, the application of detonation process in propulsion systems attracts more and more attention, resulting in several types of detonation-based engines [5–8]. Such emerging technology trend in aerospace warrants a renewed focus for detonation research in developing new technique to aid interpretation of flow data and predicting the unsteady dynamics of detonation in various combustors.

As shock-induced combustion[9–11], the detonation dynamics depends on the wave frontal structure composed of shock-flame complex. Beyond classic steady Zeldovich–von Neumann–Döring (ZND) model prediction, realistic detonation waves are featured by transverse shocks, resulting in cellular instabilities on the detonation wavefront. Understanding of the cellular structure and its evolution has long been among the most important directions in the detonation physics field. For long, the use of smoked foils to record tracks of triple points has been the standard technique in observing the cellular detonation structure. By analyzing the records, quantitative study on the cell width has been performed thoroughly and tabulated, such as its dependence of reactants, equivalence ratio, initial pressure and temperature [12]. Moreover, analytical and semi-empirical models on the cell width have been proposed, correlating this dynamic parameter with parameters determined from chemical kinetics such as the induction zone length scale [13,14].

The movement of transverse waves, which plays the key role in the cellular structure, has not been studied thoroughly. In recent years, advanced optical technologies, such as PLIF (Planar Laser Induced Fluorescence) or CTC (Computed Tomography of Chemiluminescence), provide new experimental ways to "look inside" the cellular detonation [3]. It is observed that the interaction of shock and heat release may induce various reactive front morphologies, whose regularity depends on the fuels [15]. These results reveal the dynamic structure of cellular detonation, shedding light on the detonation research beyond the cell analysis of static smoked foils. However, PLIF only provides the information of heat release zone, and the wave surface, especially location of shocks, must be measured simultaneously through schlieren photography. This is because advanced optical technologies, PLIF or CTC, are designed to capture the combustion tuned with a particular species concentration. The schlieren may be used to compensate this deficiency and get the complete wave surface, but it is limited to two-dimensional (2D) cellular detonations. On the three-dimensional (3D) detonations, CTC has shown its potential to be used to get 3D flame [16–18], but corresponding 3D shock measurement technology is still not available[19–22].

Theoretical efforts are also made to the understanding of cellular detonation dynamics and to the derivation of detonation-shock evolution equation [23], notably using the theory of detonation shock dynamics (DSD) [24–26] by simplifying the detonation shock and reaction zone with an evolving surface described by a $D_n$–$\kappa$ relationship. Using such analysis approach, worth noting are the recent works by Jackson & Short [27] and Jackson et al. [28], who apply the DSD to characterize the effect of the cellular instability on the lead shock shape and velocity evolution of the gaseous detonations, revealing some distinct features in different stages of evolution and that trajectories of all shock front portions collapsed to a common curve in velocity-curvature space. These studies provide a model to reconstruct the surface evolution and further clarify the underlying physical mechanisms for the cell motion. However, the current concepts derived from DSD suffer the underlying restriction conditions due to analytic approximations and hence, are applied only to weakly unstable detonation.

Nowadays, machine learning has become more and more ubiquitous and adopted in many scientific research disciplines. Through machine learning, computers can develop the capability to learn through training and search through data sets to predict patterns and trends. This provides a good opportunity to develop a new direction for detonation modelling. In this study, we propose a shock front reconstruction method based on the information from the heat release region. Although the parameters of post-shock heat release can be calculated theoretically given the lead detonation shock and pre-shock parameters, the reverse process, i.e., from heat release region or flame to lead shock, cannot be easily achieved to close the coupling. Benefiting from the aforementioned rapid development of machine learning, the proposed shock front reconstruction concept is based on the MLP (Multi-Layer Perceptron) modelling. MLP is found to be a powerful tool in fluid mechanics research [29–31], but its application in gaseous detonations is limited in modeling the cell width [32]. Our recent study [33] proposed a method of predicting the wave configurations of cellular detonations based on the MLP, but POD (Proper Orthogonal Decomposition) is used to extract the features of the flow fields, which is complicated and requires big data difficult to be accumulated. In this investigation, one novel reconstruction method is proposed based on only the MLP, which is trained to build up the linkage of the lead shock wavefront and the state of the reactive front. Firstly, unstable detonations obtained numerically from the reactive Euler equations are used to train the MLP and provide mapping and feedback from the heat release zone to the lead shock. The input variables for model training and the effectiveness of the proposed MLP approach for reconstructing shock front motion are discussed. To this end, it is worth noting that in principle, the proposed MLP-based shock front reconstruction method is not restricted to 1D/2D or limited by the chemical reaction model. Proper data sets, however, must be carefully provided for the MLP to be trained.

## 2. Numerical simulation methods and results

An ideal detonation model given by the reactive Euler equations with a one-step Arrhenius kinetics is considered in this work. The non-dimensional governing equations with a single-step, irreversible chemical reaction are of the form:

$$\frac{\partial \boldsymbol{U}}{\partial t} + \frac{\partial \boldsymbol{E}}{\partial x} + \frac{\partial \boldsymbol{F}}{\partial y} + \boldsymbol{S} = 0, \tag{1}$$

$$\boldsymbol{U} = \begin{bmatrix} \rho \\ \rho u \\ \rho v \\ \rho e \\ \rho \lambda \end{bmatrix}, \boldsymbol{E} = \begin{bmatrix} \rho u \\ \rho u^2 + p \\ \rho u v \\ \rho u(e+p) \\ \rho u \lambda \end{bmatrix}, \boldsymbol{F} = \begin{bmatrix} \rho v \\ \rho u v \\ \rho v^2 + p \\ \rho v(e+p) \\ \rho v \lambda \end{bmatrix}, \boldsymbol{S} = \begin{bmatrix} 0 \\ 0 \\ 0 \\ 0 \\ \dot{\omega} \end{bmatrix}, \tag{2}$$

with

$$e = \frac{p}{(\gamma - 1)\rho} + \frac{1}{2}(u^2 + v^2) - \lambda Q, \tag{3}$$

$$p = \rho T, \tag{4}$$

$$\dot{\omega} = k\rho(1-\lambda)\exp(-E_a/T). \tag{5}$$

All flow variables have been made dimensionless by reference to the uniform unburned state ahead of the detonation front,

$$\rho = \frac{\tilde{\rho}}{\tilde{\rho}_0}, p = \frac{\tilde{p}}{\tilde{p}_0}, T = \frac{\tilde{T}}{\tilde{T}_0}, u = \frac{\tilde{u}}{\sqrt{\tilde{R}\tilde{T}_0}}, v = \frac{\tilde{v}}{\sqrt{\tilde{R}\tilde{T}_0}}, Q = \frac{\tilde{Q}}{\tilde{R}\tilde{T}_0}, E_a = \frac{\tilde{E}_a}{\tilde{R}\tilde{T}_0}. \tag{6}$$

The variables $\rho$, $u$, $v$, $p$, $e$ and $Q$ are the density, velocities in $x$- and $y$- direction, pressure, total energy, and the amount of chemical heat release, respectively. For the chemical reaction, $\lambda$ is the reaction progress variable which varies between 0 (for unburned reactant) and 1 (for product). The reaction is controlled by the activation energy $E_a$ and the pre-exponential factor $k$, which is chosen to define the spatial and temporal scales, so the half-reaction zone length $L_{1/2}$, i.e. the distance required for half the reactant to be consumed in the steady ZND detonation wave, is scaled to unit length.

The governing equations are discretized on Cartesian uniform grids and solved numerically using the MUSCL-Hancock scheme with Strang's splitting. The MUSCL-Hancock scheme is formally a second-order extension to Godunov's first order upwind method by constructing the Riemann problem on the inter-cell boundary [34]. The scheme is made total variation diminishing (TVD) with the use of slope limiter MINBEE, and the Harten-Lax-van Leer-Contact (HLLC) approximate solver is used for the Riemann problem. In the simulations, we use the dimensionless parameters $Q = 50$ and $\gamma = 1.2$. These are used traditionally in numerical simulations as canonical values to investigate detonation wave phenomena [35]. The stability of the detonation is sensitive to $E_a$, which is adjusted to produce unstable detonations and cellular dynamics with different degrees of regularity.

For the 1D pulsating detonation wave simulations, an effective numerical resolution of 128 points per half-reaction length is used, which is sufficient to resolve the detailed features of the pulsating shock front with the activation energy close to the stability limit [36]. Besides, simulations allowing the detonation to run for thousands of half-reaction times are performed to ensure the ultimate correct nonlinear oscillatory behavior of the detonation propagation is achieved. The simulations are initialized by the steady solution of the ZND detonation, and zero-gradient boundary conditions extrapolated from the interior are imposed on the left and the right boundaries.

The 2D flow fields behind the cellular detonation are obtained from the simulation results of detonation wave propagating in a rectangular tube. The slip boundary conditions are used on the upper and bottom wall of the tube, while zero-gradient boundary conditions are implemented on the left and the right boundaries. Initially static unburned gas with unity density and pressure fulfills the whole tube. The ignition zone with high temperature and pressure is used to initiate the detonation, and a self-sustained detonation propagating at nearly Chapman-Jouguet velocity is formed after traveling a certain distance. Relatively lower activation energies, $E_a = 10$ and $E_a = 20$, are used to obtain cellular detonation waves. About 10 grids per $L_{1/2}$ is used for the following simulations of cellular detonations, which is shown sufficient to simulate the unstable structures. A few cases with fine grid, 20 grids per $L_{1/2}$ in the case of $E_a = 20$, are tested to see the effects of grid resolution on the reconstruction. A sufficiently large domain width of 80 is used to ensure enough detonation cells are present.

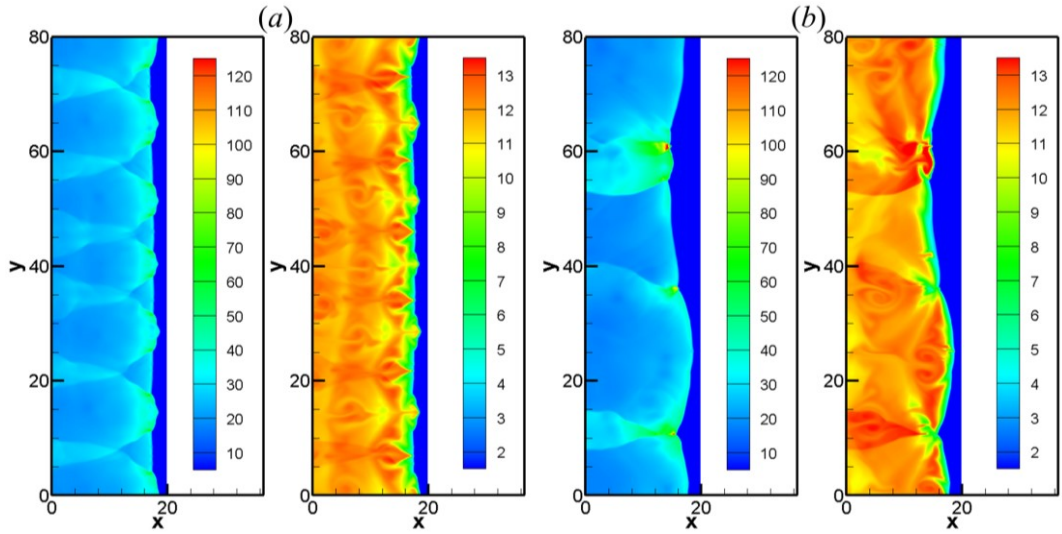

**Fig. 1** Pressure (left) and temperature (right) fields of cellular detonations with $E_a = 10$ (a) and 20 (b).

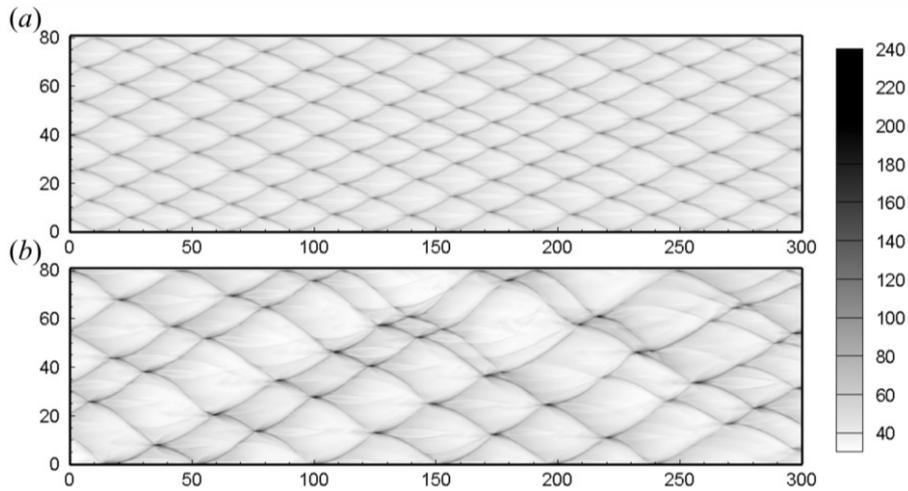

**Fig. 2** Numerical smoked foil records with $E_a = 10$ (a) and 20 (b) to illustrate the detonation cells.

The flow fields of cellular detonation after a long-time simulation avoiding the initial transient from the ignition zone are shown in Fig. 1. The self-sustained detonations are featured by cellular structures composed of reactive front, lead shock and transverse shock waves. The transverse waves propagate periodically in a direction perpendicular to the propagation of the lead shock wave. The reactive front is distorted by the lead and transverse waves, resulting in a series of irregular section. Results indicate that high $E_a$ induces more unstable detonation wave, and vice versa. With the same height, there are fewer transverse waves in the case of $E_a = 20$, and the regularity of transverse waves is weak, as shown in Fig. 1. To verify the simulated results, numerical smoked foil records using maximum pressure trace are also generated during the computation. Figure 2 shows the numerical foil records with $E_a$ = 10 and 20. Generally these results are the same as previous studies qualitatively, e.g., [35, 37–41], and can be used as samples of our shock reconstruction method.

## 3. Reconstruction method and results

### 3.1 Shock front reconstruction based on MLP

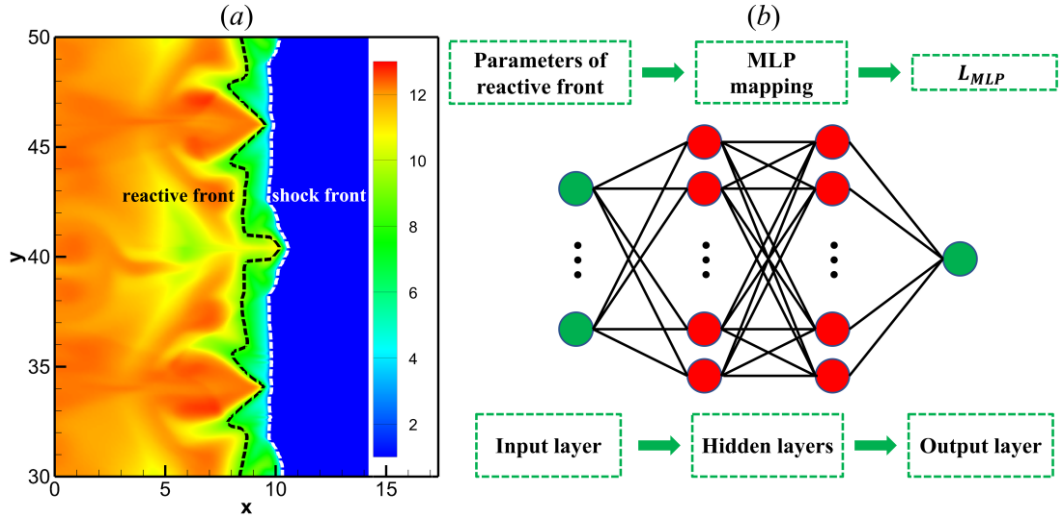

**Fig. 3** A typical temperature field (a) and basic architecture of MLP (b).

The idea of shock front reconstruction is to predict the shock motion through the information around the reactive front based on a state projection from reactive front to shock front established by MLP. In the 2D flow field as illustrated by Fig. 3(a), the reactive front is displayed by the black dashed curve, corresponding the reaction index $\lambda$ = 0.5, while the shock front by the white dashed curve, corresponding the location achieving twice of pre-shock pressure. In essence, the goal of the shock front reconstruction is to predict the location of the white dashed curve according to the flame surface parameters of the black one. Considering the parameters of reactive front are available, the problem is simplified to predict the distance $L_{MLP}$ between shock and reactive front. The use of MLP is thus to provide a nonlinear mapping relationship between $L_{MLP}$ and corresponding flame surface parameters, which can be established by learning from a large amount of shock-reactive front data extracted from the detonation wave flow field. The learning process is also known as the training process. Once the training is completed, the trained MLP can be used to quickly reconstruct complex detonation wave surface using measured or known flame surface parameters.

The basic architecture of MLP is shown in Fig. 3(b). It includes one input layer, several hidden layers, and one output layer. In input and hidden layers, there are several neurons which connect with neurons of other layers. It is a pity that there are no rigorous rules on the number of layers/neurons in the research field of MLP so far. On the problem here, several trials indicate that two hidden layers, whose neurons are fixed to be six, are enough to achieve an acceptable MLP model, thus are used in different reconstruction cases for the convenience of comparative analysis. There are no doubts that MLP can be optimized further by elaborating the MLP architecture, but this one is enough to verify our methods. The neurons of input layer are variable, changing from 3 to 6, and the output layer has only one neuron to get $L_{MLP}$. The mean of squared errors between the MLP output and the target of the training set is used here as performance function to evaluate the MLP. The activation function used in each layer plays an important role in the nonlinear mapping ability of the MLP. There are a number of common activation functions in the literature [42], and this work utilizes the hyperbolic tangent sigmoid transfer function for both two hidden layers, and the linear transfer function is used for the output layer.

In the training, we use Nguyen-Widrow layer initialization function as the MLP initial method to generate initial weight and bias values for each layer, which is useful to reduce training time [43]. The weight and bias values of neurons are updated according to Levenberg-Marquardt optimization [44,45], which is adaptive between the steepest descent method and the Gauss-Newton method to achieve fast convergency rate. When the current solution is far from the minimum, the algorithm is essentially a steepest descent method with a small step size, which is relatively slow but guaranteed converge, but gradually switches to a Gauss-Newton method to approach a quadratic approximation when the current solution is close to the correct solution. Thus, this algorithm is very efficient for training moderate-sized feedforward neural networks (up to a few hundred weights) [45]. Validation is used to stop training early if the network performance on the validation set fails to improve or remains the same for certain pre-set epochs.

Given the mixtures, many transient flow fields may exist at certain instants, like those shown in Fig. 1. For each $E_a$, a data set consisting of 60 different shock-reactive front data is generated from the transient detonation flow fields every certain number of calculation steps, after the general structure approaches a steady state. The data set is further divided randomly as the training set and validation set, with the ratio 85% and 15%. Beside the training set and validation set, the test set is generated additionally from the subsequent detonation travelling flow fields, including 10 different shock-reactive front data. For each transient flow field, we get 801 (corresponds to the grid numbers along *y*-direction) pairs of parameter values, which are extracted along the line parallel with *x*-axis. For each activation energy, one independent MLP will be trained using corresponding flow field data set described above. In order to quantitatively evaluate the reconstruction accuracy, the relative error between the MLP reconstruction length $L_{MLP}$ and the real distance between shock and reactive front $L_{real}$ is defined as follows:

$$relative\ error = \frac{L_{MLP} - L_{real}}{L_{real}} \times 100\% \tag{7}$$

There are several different choices on the input variables. Considering the conservation of mass, momentum and energy in the flow, we choose density $\rho$, temperature $T$, and velocity $u$ and $v$. The corresponding gradients of temperature and density, $T' = \partial T/\partial x$ and $\rho' = \partial \rho/\partial x$, are also introduced to verify the method, and the parameter dependence will be discussed in the later part.

## 3.2 Application on 1D pulsating detonations

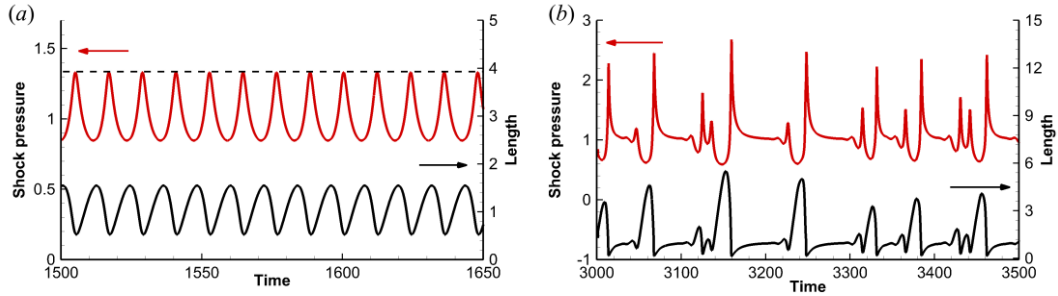

**Fig. 4** Time evolution of the lead shock pressure (red) and distance between shock and reactive front (black) for $E_a = 27.0$ (a) and $E_a = 30.0$ (b).

We first use the proposed method to reconstruct the lead shock for the 1D pulsating detonations. When the activation energy is above and close to the stability limit, the 1D detonation can exhibit oscillatory behavior with constant period. The lead shock pressure after the initiation transient passed, is plotted in Fig. 4(a) for activation energy $E_a = 27.0$. The shock pressure is normalized by the von Neumann pressure of the corresponding steady ZND solution. The time evolution of the distance between the shock and reactive front of the pulsating detonation is also shown in Fig. 4(a). As can be seen, the pulsating detonation manifests a single-mode oscillation with both lead shock pressure and shock-reactive front distance exhibiting a similar periodic trend. With $E_a$ increasing from 27.0 to 30.0, the instability of the detonation front migrates from a single-mode oscillation to a chaotic oscillation, as illustrated in Fig. 4(b). These long-time nonlinear evolution results are in good agreement with those found in literature [36,46].

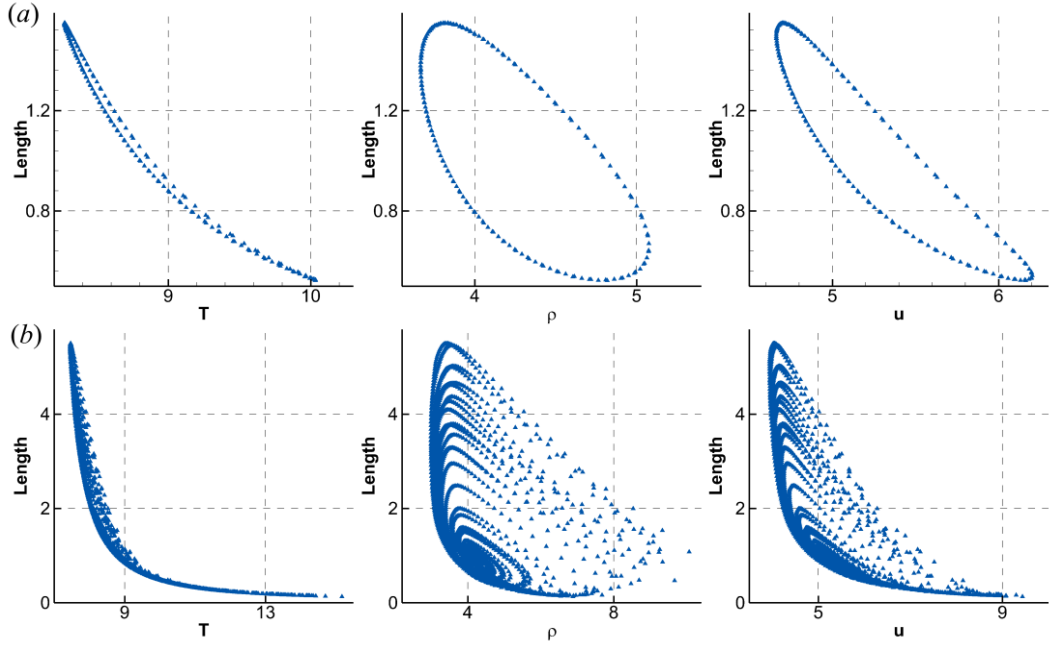

**Fig. 5** Extracted data used for training set of 1D pulsating detonations, displaying the variations of shock-reactive front distance on $T$, $\rho$, and $u$ with $E_a$ = 27.0 (a) and 30.0 (b).

Due to the different pulsating characteristic of detonations, the number of shock-reactive front data extracted for MLP training of each $E_a$ is different. For $E_a$ = 27.0, starting from the dimensionless time $t$ = 1500, we extract the training data from three consecutive oscillation cycles with the time interval $\Delta t$ = 0.2 to set up corresponding MLP. Considering that the oscillation becomes obviously irregular for $E_a$ = 30.0, more shock-reactive front data is extracted with the same time interval $\Delta t$ = 0.2 to train the corresponding MLP, i.e., from the dimensionless time $t$ = 3000 to 4000. Three input variables, i.e., $T$, $\rho$, and $u$, are used as the MLP input variables set to perform the reconstruction, since the above 1D pulsating detonations show relatively simple mapping relationships between the flame surface and the corresponding lead shock. Figure 5 displays the extracted data of each $E_a$ used for the training of each MLP, which illustrates the nonlinearity dependence between the shock-reactive front distance and three input variables. Essentially, the shock front reconstruction for 1D pulsating detonations can be regard as a multiple regression issue mathematically. In addition to MLP, there are many other widely used traditional regression models. Considering the interpretability and the explicitness of different models, we utilize multiple linear regression (MLR) and multivariate second-degree polynomial regression (MSPR) to reconstruct the shock front as comparisons to the proposed

MLP model.

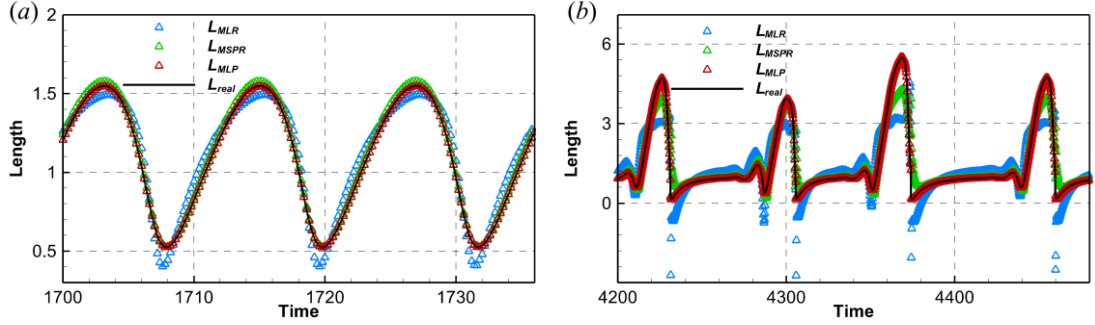

**Fig. 6** Reconstruction results of $E_a$ = 27.0 (a) and 30.0 (b) with different models, reconstruction input variables set $T$, $\rho$, and $u$.

**Table 1** Average relative error of 1D pulsating detonations with different reconstruction models.

| Reconstruction models | $E_a$ = 27.0 ($\Delta t$ = 0.2) | $E_a$ = 27.0 ($\Delta t$ = 0.1) | $E_a$ = 30.0 ($\Delta t$ = 0.2) | $E_a$ = 30.0 ($\Delta t$ = 0.1) |
|---|---|---|---|---|
| MLR | 7.293% | 7.275% | 53.394% | 53.101% |
| MSPR | 2.023% | 2.012% | 32.535% | 30.715% |
| MLP | 0.085% | 0.082% | 0.314% | 0.313% |

The three different well-established models are used to reconstruct the shock-reactive front distance of the 1D pulsating detonations from the dimensionless time $t$ = 1700 and 4200 for $E_a$ = 27.0 and 30.0, respectively. The reconstruction results corresponding to different activation energies, in which the test data is extracted with time interval $\Delta t$ = 0.2, are shown in Fig. 6. And Table 1 gives the average relative errors of each $E_a$ test set with three different models. It can be seen that the prediction results of the three models are in good agreement with the real shock-reactive front distance curve of $E_a$ = 27.0. However, as the $E_a$ increases to be 30.0, the reconstructing errors increase significantly for MLR and MSPR. The prediction results of the two models have large deviations when the pulsating detonation exhibits violent oscillation at the peak and trough of each pulsation cycle. Especially for MLR, there are even non-physical results with negative lengths shown in Fig. 6(b). On the contrary, MLPs maintain good performance for different activation energies with all the average value of relative error below 0.5%, which indicates that MLP has a better nonlinear learning and prediction ability to reconstruct shock front of the 1D pulsating detonations. Although using the MLP introduces a "black-box" to reconstruct shock front, these comparison results of different models demonstrate a tradeoff between the interpretability and the prediction accuracy. Furthermore, for 2D or even 3D cellular detonations, MLP is able to take full advantage of its powerful big data learning ability from the prospective of better reconstruction and generalization. Furthermore, reconstruction results of average relative error for test data extracted with time interval $\Delta t = 0.1$ are also shown in Table 1. It is found out that the errors corresponding different time interval are close with the same $E_a$ and reconstruction model, demonstrating that the time interval $\Delta t$ of extracting test set data has little influence on the reconstruction results.

### 3.3 Application on 2D cellular detonations

Compared with the 1D pulsating detonations, the additional $y$-direction in the 2D flow field brings cellular surface of detonation waves along with reciprocating transverse shock waves, which makes the shock surface difficult to reconstruct. Consequently, more information from reactive front need to be provided to achieve high quality reconstruction. Thus, six parameters, $T$, $T'$, $\rho$, $\rho'$, $u$ and $v$, are first used as MLP input variables set of the flow field reconstruction method mentioned above to reconstruct the lead shock surface of the 2D unstable cellular detonations. Detailed discussion on MLP input variables of 2D reconstruction will be presented in following Sec 3.4.

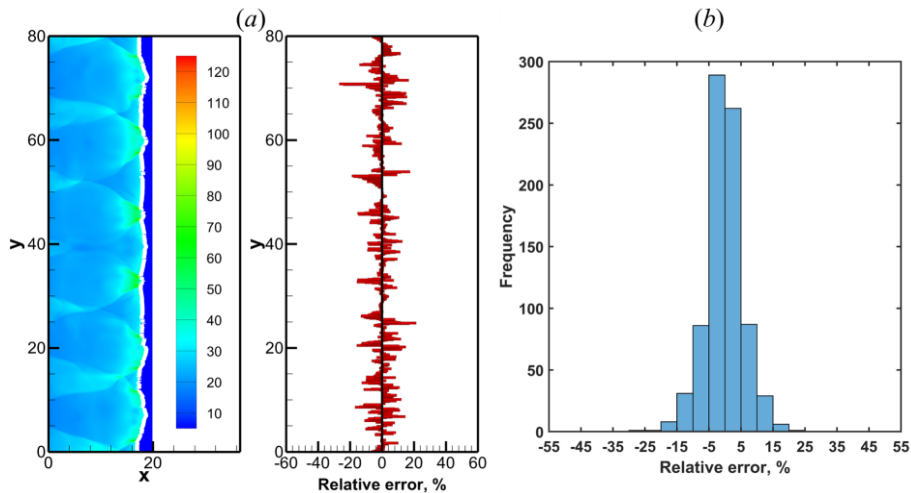

**Fig. 7** Typical reconstructed shock (white curve in pressure field) and relative error of shock distance with $E_a = 10$ (10 grids per $L_{1/2}$) (a) and corresponding relative error frequency distribution histogram (b), MLP input variables set $T$, $T'$, $\rho$, $\rho'$, $u$ and $v$.

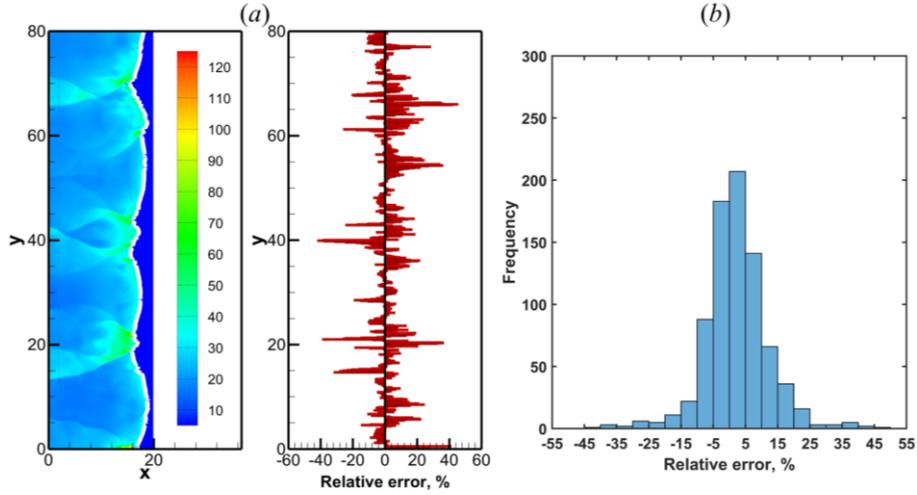

**Fig. 8** Typical reconstructed shock (white curve in pressure field) and relative error of shock distance with $E_a = 20$ (10 grids per $L_{1/2}$) (a) and corresponding relative error frequency distribution histogram (b), MLP input variables set $T, T', \rho, \rho', u$ and $v$.

Typical reconstruction results from test set for the cases of $E_a = 10$ and 20 are shown in Figs. 7(a) and 8(a), respectively. It is observed that the reconstructed shock, plotted by the white curve in the flow fields, locates around the simulated lead shock, demonstrating that a well-trained MLP can predict the shock influenced by the reactive front precisely. The relative errors at different positions are also listed quantitatively. The error may be positive or negative, meaning the distance $L_{MLP}$ may be larger or smaller than its real value, but in a limited range. The error range is below 20% in the case of $E_a = 10$, but becomes large in the case of $E_a = 20$.

The distribution of relative error is shown in Figs. 7(b) and 8(b) to facilitate further discussion. In the case of $E_a = 10$, close to 600 results among 801 pairs of data have an absolute value of relative error less than 5%. Although the error range becomes large when $E_a$ increases to 20, again about half of the results have an absolute value of relative error less than 5%, which still supports the reasonable performance of this MLP-based reconstruction approach. In fact, it is not surprising that increasing $E_a$ results in larger error. Results of both flow fields in Fig. 1 and numerical smoked foils in Fig. 2 demonstrate that the detonation becomes unstable when increases $E_a$. This makes the relation of the shock and reactive front more involved, so raises the difficulty to reconstruct the shock. As can be seen from the MLP reconstruction results for 1D pulsating detonations, the results are almost the same as the simulated ones. In 2D cellular detonations, the leading shock is curved and with discontinuities induced by transverse shocks, these lead to unavoidable sources of error.

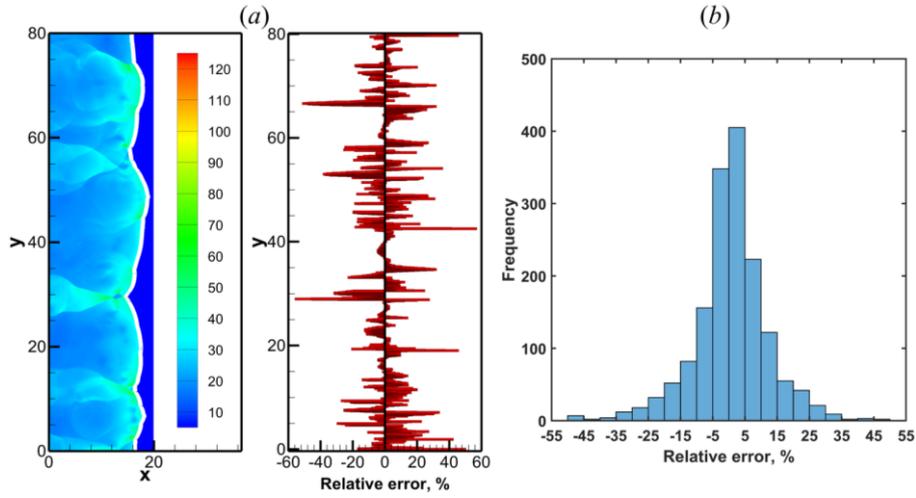

**Fig. 9** Typical reconstructed shock (white curve in pressure field) and relative error of shock distance with $E_a = 20$ (20 grids per $L_{1/2}$) (a) and corresponding relative error frequency distribution histogram (b), MLP input variables set $T, T', \rho, \rho', u$ and $v$.

To verify the influence of grid scale on the reconstruction results of MLP, 20 grids per $L_{1/2}$ is used to carry out simulations for $E_a = 20$. The same procedure and MLP parameters are used to train and test the MLP, except that 801×2 pairs of shock-reactive front data are extracted from each transient flow field. For the cases of six input variables, i.e., $T$, $T'$, $\rho$, $\rho'$, $u$ and $v$, typical reconstruction results from test set of $E_a = 20$ with fine grid are shown in Fig. 9. Results show that trained MLP still can reproduce the shock based on the parameters of the reactive front accurately. Figure 7(b) shows the corresponding error frequency distribution histogram, which has a similar distribution characteristic to that of Fig. 8(b) for $E_a = 20$ with 10 grids per $L_{1/2}$.

### 3.4 Discussion on input variables and generalization

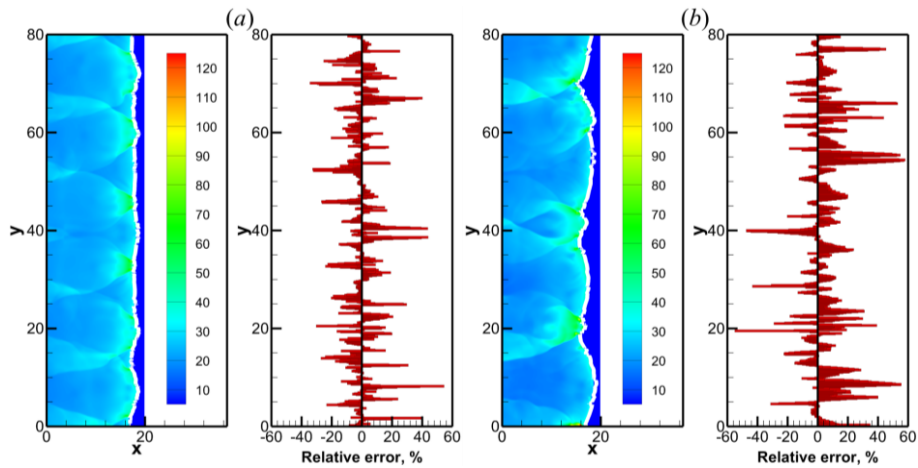

**Fig. 10** Typical reconstructed shock (white curve in the pressure fields) and relative error of shock

distance with $E_a$ = 10 (a) and 20 (b), MLP input variables set $T$, $\rho$, $u$ and $v$.

Results in Sec. 3.2 demonstrate that the idea of using MLP to reconstruct the lead shock of both 1D and 2D unstable detonations is feasible primarily with 6 input variables. Since the flame surface information obtained from experiment results are usually very limited for most of practical cases, whether the proposed method is still effectiveness when the number of input variables for MLP is reduced, is especially vital to its engineering application. Thus, more reconstruction cases with less input variables are carried out to test the practicability of the proposed method. These tests are performed by using 4 input variables ($T$, $\rho$, $u$ and $v$), and reconstruction results from test set for the cases of $E_a$ = 10 and 20 cellular detonation flow fields are shown in Fig. 10. Generally speaking, the error increases noticeably after the gradients of temperature and density are removed. From the flow fields shown in Fig. 10, the reconstructed shock, displayed by white curve, remains nevertheless close to the simulated one. Additionally, the high activation energy of $E_a$ = 20 induces again large error, due to the strong instability of corresponding 2D cellular detonations, which is similar to the results with six input variables.

**Table 2** Reconstruction results of test set with different MLP input variables set.

| Input variables set | $E_a$ = 10 (10 grids per $L_{1/2}$) | $E_a$ = 20 (10 grids per $L_{1/2}$) | $E_a$ = 20 (20 grids per $L_{1/2}$) |
|---|---|---|---|
| $T$, $T'$, $\rho$, $\rho'$, $u$, $v$ | 4.10% | 7.04% | 7.33% |
| $T$, $\rho$, $u$, $v$ | 7.46% | 11.10% | 9.77% |
| $T$, $T'$, $u$, $v$ | 9.75% | 14.77% | 13.93% |
| $\rho$, $\rho'$, $u$, $v$ | 4.86% | 8.28% | 8.00% |
| $T$, $T'$, $\rho$, $\rho'$ | 8.43% | 11.41% | 12.57% |

To give an overall estimation of reconstruction errors, the average relative error of all reconstruction test samples is calculated. Each input variables set has been trained for five times to exclude the effects of initiation in MLP, whose average error is shown in Table 2. Besides the two groups of input variables, three other groups with different input variables were performed. In the given five groups, the first group gives the best results, and removing the gradients of temperature and density increases the error from 4.10% to 7.46% with $E_a$ = 10 and from 7.04% to 11.10% with $E_a$ = 20. The other groups remove the input variable of density, temperature and velocity, respectively. It is found out that the 4th group gives the best results when keeping 4 input variables, in which the related variables of temperature are removed. We deduce that in the flow fields, the temperature variation between the shock and reactive front is modest, so plays a relatively weaker role in the reconstruction. Furthermore, effects of resolution are also tested through the $E_a = 20$ cases and shown in Table 2. It is found out that the errors corresponding the same input variables set are close with the same $E_a$, demonstrating that the reconstruction is not significantly sensitive to the accuracy of simulated results. Further analysis on the role of different variables, as well as the choosing strategy, is necessary in the future.

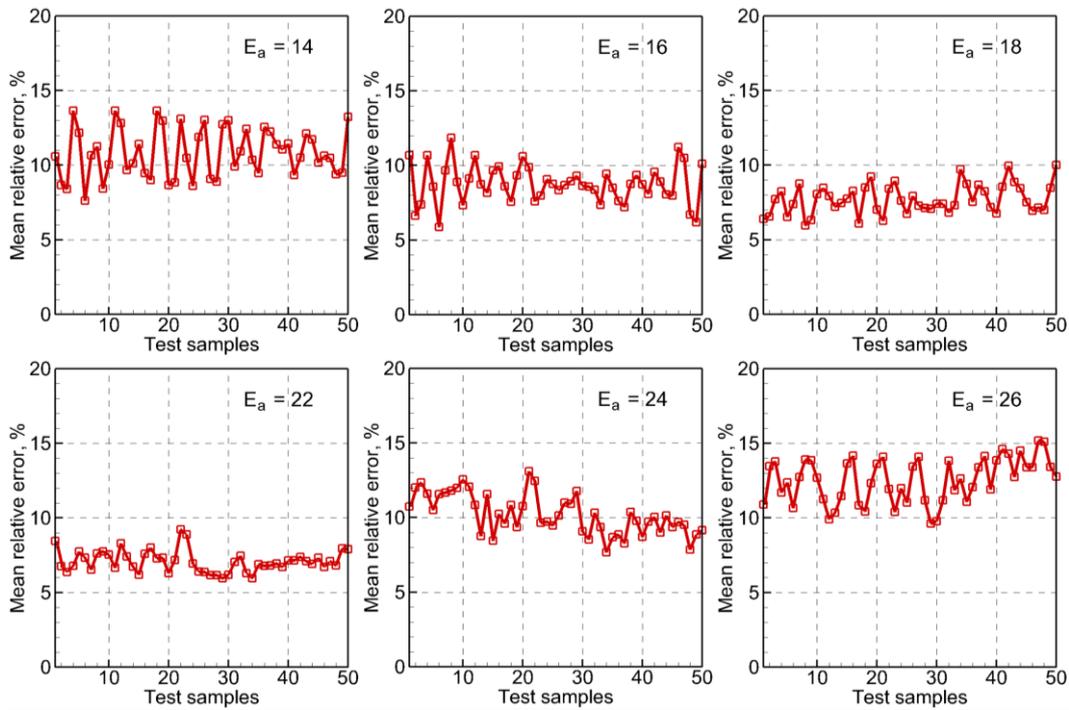

**Fig. 11** Error distribution of reconstruction results of six different $E_a$ test set

Another key problem of shock front reconstruction based on MLP is the generalization capability. It has been demonstrated that a well-trained MLP can accurately reconstruct the lead shock wavefront, given a fixed activation energy $E_a$. Considering the activation energy may vary in practical applications, further tests should be performed to examine the generalization capability at different activation energies. If the proposed method is robust in the lead shock reconstruction for different activation energies, it will be a flexible and powerful tool for future application on shock front reconstruction for detailed chemical reactions simulation and experimental results. In these further tests, we use the well-trained MLP from the case of $E_a = 20$ with six input variables ($T$, $T'$, $\rho$, $\rho'$, $u$, $v$), and the cellular detonation flow fields with six other activation energies, $E_a = 14, 16, 18, 22, 24, 26$, are reconstructed.

The main parameters of numerical simulations are the same, such as domain width 80 and 10 grids per $L_{1/2}$, and the test set for each $E_a$ is composed of 50 different transient flow fields extracted from different instants with every certain number of calculation steps. Figure 11 shows the mean relative error of every flow field for each $E_a$ test set. For each flow field, the error is given by a point. Generally speaking, the relative error increases when $E_a$ deviates from 20, the activation energy used to train the MLP. In the cases of $E_a$ = 18 and 22, all the error points locate below 10%. In contrast, the error points mainly locate between 10% and 15% in the cases of $E_a$ = 14 and 26. This should be attributed to the different mapping relationships between the reaction surface and the corresponding lead shock induced by the activation energy variation.

**Table 3** Average errors based on 50 flow fields in the cases of each $E_a$.

| $E_a$ | Average error |
|---|---|
| 14 | 10.83% |
| 16 | 8.77% |
| 18 | 7.72% |
| 20 | 7.04% |
| 22 | 7.09% |
| 24 | 10.21% |
| 26 | 12.54% |

Although the error increases when $E_a$ either increases or decreases, we think that the deviation is acceptable, and the average errors of all flow fields are shown in Table 3. It should be noted that the error has reached above 7% already in the case of $E_a$ =20 with the test samples generated from the same $E_a$ flow fields. When $E_a$ decreases to be 14, the error increases to be about 10.83%, less than twice of the basic error. However, it is surprised that the error does not increase dramatically when $E_a$ increases. When $E_a$ increases to be 26, the error is about 12.54%, still less than twice of the basic error. Theoretically, increasing $E_a$ introduces a number of unstable modes, and then leads to the more complicated cellular structures, as illustrated in Fig. 2. Nevertheless, the MLP trained by low $E_a$ data still performs well given high $E_a$, indicating the proposed method has a good generalization capability.

# 4. Conclusion

A novel method of reconstructing the lead shock of unstable detonation is proposed and tested in this study. Within the detonation structure, the lead shock and the reactive front are coupled. Benefiting from the rapid development of advanced laser-based optical diagnostics within the combustion community, measurement of the flow and kinetic details as well as the location of reactive front can be achieved. It is the objective of this work to link such information to predict the lead shock evolution and to obtain a complete picture of cellular detonation dynamics. Using multi-layer perceptron (MLP) method, we propose a shock reconstruction method which can predict the lead shock evolution from details of the reactive front. The application of the proposed method is analyzed and verified thoroughly through the numerical results of 1D pulsating detonations and 2D cellular detonations using the reactive Euler equations with a one-step irreversible chemical reaction model. For detonations with two activation energy values, effects of input variables number are also studied by analyzing the error of trained MLP. Furthermore, the extensible of reconstruction method is investigated by reconstructing flow field with different activation energies, which shows the proposed method has well generalization capability.

This work is performed based on 2D cellular detonation from numerical simulations, and one-step irreversible chemical reaction model is employed to generate the data set. However, it should be noted that this shock reconstruction method is universal and expandable. It should be not limited to 2D, but capable of reconstructing 3D shock. The 1D pulsating detonations reconstruction results show that the average relative error is below 1%. From 1D to 2D, the curved shock may raise the error, so it is expected of larger error from 2D to 3D, but there should be no principal obstructions. On the other side, the one-step irreversible chemical reaction model is actually a limitation to this method. With detailed chemical reactions, more species distributions are simulated and thus defining the reactive front has more choices, helpful to improve the MLP reconstruction. Moreover, the present concept may provide a novel way of combing numerical and experimental results. The numerical simulations can generate big data, in which the projection of many different characteristic values can be extracted. In contrast, the lack of enough information in experiments becomes serious, such as the reactive front without lead shock measured at the same time. The MLP can thus be used widely to combine the numerical and experimental results, and provide a new framework to interpret results of detonation waves.

As a preliminary work, a rather simple MLP is chosen to focus the feasibility of the reconstruction idea. Undoubtedly, the results will be improved greatly benefiting from more advanced deep learning technologies developed recently. One of the main problems of this method derives from the MLP output, with only one neuron predicting the shock and reactive front distance $L_{MLP}$. Theoretically, this mapping should be performed along the streamlines, so this method performs well in 1D detonations. For 2D detonations, the reactive front not only influences the lead shock exactly ahead, but also the neighboring shock. Therefore, the one neuron is oversimplified, and the scalar $L_{MLP}$ should be replaced by a vector. Furthermore, using the information from historic flow fields is also a good idea considering the pattern of cellular detonations. These spatial and temporal physical consideration could be implemented by the new technologies in the machining learning, such as CNN (Convolution Neural Network) or RNN (Recurrent Neural Network).

## Acknowledgements

The research is supported by the National Natural Science Foundation of China NSFC (No. 11822202)